\documentclass[twocolappendix]{emulateapj}

\usepackage[usenames,dvipsnames]{color}

\usepackage{comment}

\usepackage{soul,color,amsmath}

\shorttitle{Nonlinear Corotation Torques}
\shortauthors{P.~C.~Duffell}

\begin{document}

\author{Paul C.~Duffell}
\affil{Theoretical Astrophysics Center, University of California, Berkeley}
\email{duffell@berkeley.edu}

\title{Halting Migration:  Numerical calculations of corotation torques in the weakly nonlinear regime}

\begin{abstract}

Planets in their formative years can migrate due to the influence of gravitational torques in the protoplanetary disk they inhabit.  For low-mass planets in an isothermal disk, it is known that there is a strong negative torque on the planet due to its linear perturbation to the disk, causing fast inward migration.  The current investigation demonstrates that in these same isothermal disks, for intermediate-mass planets, there is a strong positive nonlinear corotation torque due to the effects of gas being pulled through a gap on horseshoe orbits.  For intermediate-mass planets, this positive torque can partially or completely cancel the linear (Type I) torque, leading to slower or outward migration, even in an isothermal disk.  The effect is most significant for Super-Earth and Sub-Jovian planets, during the transition from a low-mass linear perturber to a non-linear gap-opening planet, when the planet has opened a so-called ``partial gap", though the precise values of these transition masses depend sensitively on the disk model (density profile, viscosity, and disk aspect ratio).  In this study, numerical calculations of planet-disk interactions calculate these torques explicitly, and scalings are empirically constructed for migration rates in this weakly nonlinear regime.  These results find outward migration is possible for planets with masses in the $20 - 100 M_{\earth}$ range, though this range depends on the disk model considered.  In the disk models where torque reversal occurs, the critical planet-to-star mass ratio for torque reversal was found to have the robust scaling $q_{\rm crit} \propto \sqrt{\alpha}(h/r)^3$, where $\alpha$ is the dimensionless viscosity parameter and $h/r$ is the disk aspect ratio.

\end{abstract}


\section{Introduction}
\label{sec:intro}

Planetary migration is a pivotal process in the formation of planetary systems \citep{2012ARAnA..50..211K}.  During the formation of the planets, gravitational torques are exerted on these satellites by the protoplanetary disk they inhabit which can enable substantial migration during the disk's lifetime.

Analytical estimates of the migration rate usually focus on either the linear ($m_p \sim$ Earth mass) regime, or the strongly nonlinear ($m_p \gtrsim$ Jupiter mass) ``gap-opening" regime.  The linear case (Type I migration) is straightforward to calculate by summing over Lindblad resonances \citep{1980ApJ...241..425G, 1986Icar...67..164W}.  For this regime, it has been shown that fast, inward migration is the destiny of planets in an isothermal disk, although more recently it has been shown that relaxing the equation of state can change the direction of migration \citep{2006AnA...459L..17P}.  In the nonlinear case, the dynamics can be much more complicated (for instance a gap may form), and therefore some assumptions must be made in order to gather any estimate for the migration rate.  Often, one assumes that the gap acts as a dam separating the disk into an inner and outer region.  If this is the case, and assuming the planet's inertia is negligible, the planet would have to migrate at the viscous drift rate of the disk.  This is known as Type II migration \citep{1986ApJ...309..846L, 1997Icar..126..261W}.

These estimates must of course be verified by experiments, and at the moment the only experimental test-beds for planet migration are numerical simulations.  The Type I theory has been thoroughly validated by these experiments \citep{2002ApJ...565.1257T, 2010ApJ...724..730D}, but given the assumptions that go into the Type II prediction, surprisingly few systematic parameter-space studies have been devoted to explicitly validate this theory \citep[but see][]{2007ApJ...663.1325E, 2008arXiv0807.0625E}.  In fact, the idea that gap-opening ties migration to the viscous rate has been called into question; in recent studies, it has been shown that Jupiter-mass planets with deep gaps can migrate at rates much slower or faster than the viscous rate, contrary to the Type II prediction \citep{2014ApJ...792L..10D, 2014arXiv1411.3190D}.  This is due to the fact that the planet does not truly act as a ``dam" in multi-dimensional models, a fact which has been affirmed in many studies \citep{1999ApJ...526.1001L, 1999MNRAS.303..696K, 2001MNRAS.320L..55M, 2003MNRAS.341..213B, 2006ApJ...641..526L, 2007AnA...461.1173C, 2014ApJ...782...88F}.  Moreover, only a few studies have systematically explored the parameter space of the ``partial gap" regime, between linear and nonlinear theory \citep{2003MNRAS.341..213B, 2006ApJ...652..730M, 2007MNRAS.377.1324C, 2011MNRAS.410..293P}.

The present study focuses on this transition from the linear to the nonlinear regime, when the planet mass is a few tens that of Earth.  This regime is particularly pertinent because linear theory predicts very fast migration rates for such planets, a prediction which may be in tension with the abundance of Super-Earths which survive to be in exoplanet surveys like Kepler \citep{2008ApJ...673..487I, 2009AnA...501.1139M, 2009AnA...501.1161M}.

In this work, it is found numerically that in the intermediate-mass regime the partial gap can exert a significant outward torque on the planet, strong enough to reduce the migration rate substantially and, for some planets, reverse the direction of migration.  This outward torque can be straightforwardly understood as a nonlinear corotation torque due to material being pulled from the outer disk and quickly being deposited on the inner disk after executing a horseshoe orbit, similar to the interpretation invoked by \cite{2003ApJ...588..494M} to describe the torques responsible for Type III migration.  Note that strong outward corotation torques have previously been seen for very low-mass planets in non-barotropic disks \citep{2006AnA...459L..17P, 2011MNRAS.410..293P, 2013AnA...549A.124B}.  The strong outward torques in this study are from intermediate-mass planets in isothermal disks.  The condition of isothermality is applicable at large radii ($> 10$ AU), where the orbital time is longer than the thermal relaxation time of the disk \citep{2013AnA...549A.124B}.

Torques produced by linear theory have the following analytical scaling with disk and planet parameters:

\begin{equation}
\Gamma_0 \equiv q^2 \mathcal{M}^2 \Sigma_0 a^4 \Omega_p^2,
\end{equation}

where $q$ is the planet-to-star mass ratio, $\mathcal{M}$ is the Mach number (equivalently the inverse of the disk aspect ratio $\mathcal{M} = (h/r)^{-1}$), $\Sigma_0$ is the unperturbed surface density at the planet's position, $a$ is the radius of the planet's orbit, and $\Omega_p$ is the Keplerian orbital frequency at the planet's position.  The linear torque on the planet should be of order this characteristic torque:

\begin{equation}
\Gamma_{\rm linear} = - A \Gamma_0,
\label{eqn:lintorques}
\end{equation}

where $A$ is some dimensionless order-unity constant, independent of $q$, $\alpha$ or $\mathcal{M}$.  Given this, the planet's migration rate can be calculated straightforwardly as $\dot a / a = 2 \Gamma / a^2 \Omega_p m_p$ where $m_p$ is the planet mass, so

\begin{equation}
\dot a / a = - 2 A q \mathcal{M}^2 {\Sigma_0 a^2 \over M_*} \Omega_p.
\end{equation}

where $M_*$ is the stellar mass.  Numerical calculations in the low-mass limit \citep[e.g.][]{2002ApJ...565.1257T} have shown that this constant $A$ is generally positive for all isothermal disk models, guaranteeing consistent inward migration for such disks.

It has been argued \citep{2008AnA...485..877P} that corotation torques have a fundamentally nonlinear nature, and therefore predictions from linear theory give an incomplete picture.  Linear corotation torques scale identically to the Lindblad torques, and therefore are degenerate with them, representing an overall correction to the constant $A$ in (\ref{eqn:lintorques}).  \cite{1991LPI....22.1463W} suggested a nonlinear formula:

\begin{equation}
\Gamma \propto x_s^4 \Sigma_0 \Omega_p,
\end{equation}

where $x_s$ is the gap width.  However, after substitutions, assuming $x_s \sim a \sqrt{q \mathcal{M}}$ \citep{2002AnA...387..605M, 2006ApJ...652..730M}, this formula also scales identically to the Lindblad torque and is therefore also degenerate with it.  For isothermal disk models, this produces only modest corrections to the constant $A$, leaving the sign fixed, and ensuring consistent inward migration.

\cite{2011MNRAS.410..293P} developed a model for intermediate-mass planets (for both barotropic and non-barotropic disks), calibrated using highly accurate numerical calculations (although the parameter space was not explored in detail).  These results suggested that strong nonlinear corotation torques may be possible for these planets, but torque reversal was not reported for isothermal disks in that study.  Torque reversal was, however, seen in the thorough parameter study by \cite{2006ApJ...652..730M}.

The strength of the corotation torque might be expected to be modified due to the presence of a partial gap.  Suppose a partial gap is formed, so that the density is somewhat lower in the corotation region, but a steady mass flux is still present across the gap on horseshoe orbits.  This implies that the drift velocity in the corotation region is larger than the viscous drift rate.  A fluid element viscously drifts inward until reaching the outer edge of the gap, at which point it is pulled quickly from the outer disk and deposited onto the inner disk, after which it continues to viscously drift inward.  This process of moving a fluid element from the outer disk to the inner disk removes angular momentum from the fluid element, and this angular momentum can be imparted onto the planet.  The partial gap is important because the average drift velocity can be substantially larger in the partial gap than in the rest of the disk, suggesting the planet is exerting additional torque on fluid elements in the partial gap.  This general idea has previously been invoked by \cite{2003ApJ...588..494M} to explain the torques responsible for Type III migration, and is described in much more detail there.

For the most part, this study will remain agnostic to the detailed interpretation of these nonlinear torques.  Instead of building a model to predict the appropriate scalings of the nonlinear torques, these scalings will be determined empirically through direct numerical calculations, regardless of the underlying interpretation.

Since the formula (\ref{eqn:lintorques}) assumes only linear perturbations to the disk, there are of course higher-order corrections to this formula:

\begin{equation}
\Gamma = - A \Gamma_0 + O(q^3).
\end{equation}

This work will search for the next-order term:

\begin{equation}
\Gamma = \Gamma_0( -A + B q/q_2 ) + O(q^4),
\label{eqn:six}
\end{equation}

where $B$ is some new dimensionless constant, and $q_2$ is a characteristic mass ratio of the nonlinear corotation torque, which will depend on the disk parameters, $q_2 = \alpha^a \mathcal{M}^b$ (The semi-analytical predictions of \cite{2011MNRAS.410..293P} suggest the scaling $q_2 = \alpha^{2/3} / \mathcal{M}^{7/3}$.  However, it will be found in this study that $q_2$ robustly scales as $q_2 = \sqrt{\alpha}/\mathcal{M}^3$).  This scaling and the magnitude of the constant $B$ will determine the strength of nonlinear corotation torques, and whether these torques are strong enough to cause a reversal of the direction of migration.

In particular, this study defines the nonlinear torque as

\begin{equation}
\Gamma_2 \equiv \Gamma - \Gamma_{\rm linear},
\end{equation}

and the normalized nonlinear torque as

\begin{equation}
t(q) \equiv \Gamma_2 / \Gamma_0 = B q/q_2 + O(q^2).
\label{eqn:tq}
\end{equation}

The scaling of $q_2$ and the value of the constant $B$ will be found by numerically exploring the parameter space of $q$, $\alpha$ and $\mathcal{M}$ for a large number of planet-disk systems.

\begin{table} \caption{List of Variables and Constants} \label{tab:engine}
\begin{center} \leavevmode \begin{tabular}{lll} \hline \hline Variable  
    & Definition              & Formula          \\ 
\hline 
 $a$            & Orbital Radius              &  \\
 $\Omega_p$     & Planet Orbital Frequency    &  \\
 $\Sigma_0$     & Unperturbed Surface Density &  \\ 
 $q$            & Mass Ratio                  &  \\
 $\mathcal{M}$  & Mach Number                 &  \\
 $c$            & Sound Speed                 & $a \Omega_p / \mathcal{M}$ \\
 $h$            & Disk Scale Height           & $a / \mathcal{M}$ \\
 $\nu$          & Kinematic Viscosity         &  \\ 
 $\alpha$       & Dimensionless Viscosity Parameter & $\nu / (h c) $ \\
 $\Gamma_0$     & Linear Torque Scaling       & $q^2 \mathcal{M}^2 \Sigma_0 a^4 \Omega_p^2$ \\
 $\Gamma_2$     & Nonlinear Torque            & \\
 $t(q)$         & Normalized Nonlinear Torque & $\Gamma_2 / \Gamma_0$ \\
 $\dot M$       & Accretion Rate              & $(3/2) \nu \Sigma_0$\\ 
 $q_{2}$        & Characteristic Mass Ratio   & $\alpha^{1/2} / \mathcal{M}^3$ \\ 
 $q_{NL}$       & Nonlinear Mass Ratio        & $1 / \mathcal{M}^3$ \\ 
\hline \multicolumn{3}{l}{}                                            
\\ \end{tabular} \end{center} \end{table}

\section{Numerical Method}
\label{sec:numerics}

The calculations used in this study are a numerical integration of the 2D isothermal fluid equations:

\begin{eqnarray}
\partial_t \Sigma + \nabla \cdot ( \Sigma \vec v ) ~& = &~ 0 
\label{eqn:cons1}
\\
\partial_t ( \Sigma v_j ) + \partial_i ( \Sigma \vec v_i v_j + P \delta_{ij} + \sigma_{ij} ) ~& = &~ -\Sigma \partial_i \phi ~~~~~
\label{eqn:cons2}
\end{eqnarray}

\begin{equation}
P = c^2 \Sigma,
\end{equation}

where $\Sigma$ is surface density, $\vec v$ is velocity, $P$ is pressure, $c$ is the sound speed, and $\phi$ is the gravitational potential of the two bodies.  The sound speed (or equivalently temperature) as a function of radius is prescribed by hand.  For this work, the entire disk is at a single fixed temperature, so that the sound speed is fixed as 

\begin{equation}
c = a \Omega_p / \mathcal{M},
\end{equation}

where $a$ is the orbital radius of the planet, and $\Omega_p$ is its orbital frequency.  The Mach number $\mathcal{M}$ is a freely chosen parameter (equivalent to varying the disk aspect ratio, as $h/r = \mathcal{M}^{-1}$).  The $\sigma_{ij}$ in (\ref{eqn:cons2}) is the viscous stress tensor, proportional to $\nu$, the kinematic viscosity.  Most of the tests presented here assumed a uniform viscosity throughout the disk (rather than a Shakura-Sunyaev $\alpha$ viscosity), for reasons of simplicity and ease of interpretation.  However, results will still be reported in terms of the dimensionless viscosity $\alpha = \nu / hc$.

Numerical calculations are carried out using the highly accurate DISCO code \citep{2012ApJ...755....7D, 2013ApJ...769...41D}, a moving-mesh hydrodynamics code tailored specifically to the study of gaseous disks.  Calculations used 512 radial zones, for a resolution of $h/\Delta r \sim 10$ zones per scale height in the vicinity of the planet.  It has already been demonstrated in high-resolution code tests \citep{2013ApJ...769...41D} that such a resolution gives an effective numerical viscosity of $\alpha_{\text{num}} \sim 3 \times 10^{-5}$, well below any explicit viscosity in the parameter space of this study.  The azimuthal resolution varies with radius, as it is chosen to give zones with aspect ratio very close to unity, $r \Delta \phi = \Delta r$.  Therefore, at the planet's position, the azimuthal resolution is 1397 zones.

Each system was evolved for 5000 orbits (roughly a viscous time).  For many (but not all) of these systems, such long timescales were necessary in order to achieve a steady-state solution.

\subsection{Initial and Boundary Conditions}

In order to keep the results simple and easy to interpret, a very simplified isothermal disk model is assumed with

\begin{equation}
\Sigma(r) = \Sigma_0
\end{equation}

\begin{equation}
\Omega(r) = \Omega_p (a/r)^{3/2}
\end{equation}

\begin{equation}
v_r(r) = -\frac32 \nu / r.
\end{equation}

This specifies the initial conditions (because of scale invariance, $\Sigma_0$ is arbitrary).  The boundaries at $r_{min} = 0.25$ and $r_{max} = 2.5$ are Dirichlet, i.e. they are fixed to their initial values, ensuring a constant mass flux through the system.  This disk profile is a steady-state solution to the unperturbed hydro equations.  No reflections were observed from these boundaries at late times, and confidence in this scheme can be bolstered by the extremely accurate reproduction of linear theory (Figure \ref{fig:tdens}).

It must be noted here that this disk model may be expected to produce particularly large corotation torques due to the large gradient in vortensity present in the disk, when compared with a disk model in which $\Sigma \propto r^{-3/2}$, such as the Minimum Mass Solar Nebula (MMSN).  The reason that the uniform disk model is chosen here is that it is much easier to pick out the corotation torques and to measure their scalings.  In order to improve the completeness of this work, some additional calculations are also performed with disk models in which $\Sigma \propto r^{-3/2}$.  This disk also uses a viscosity $\nu \propto r^{3/2}$ corresponding to an $\alpha$ disk, so that the background flow is still a steady-state inward accretion.  The torque in these calculations will be found to have the same basic scalings and signs, but with different overall dimensionless constants in front.  Since it is unknown how closely protoplanetary disks agree with the strict MMSN scaling, uniform disk models (or at least disks with scalings shallower than $r^{-3/2}$) are still potentially relevant cases.

\subsection{Planetary Potential}

The potential $\phi(\vec x)$ in Equation (\ref{eqn:cons2}) is given by

\begin{equation}
\phi(\vec x) = - G M_{*} \left( { 1 \over | \vec x | } + { q \over \sqrt{ (\vec x - \vec x_p)^2 + \epsilon^2 } } \right).
\end{equation}

In order to simplify the problem so that it can be studied as simply and systematically as possible, this study assumes that the primary is fixed and that the planet's drift velocity is negligible, i.e. the gas is a test fluid.  The divergent planetary potential is smoothed over the length scale $\epsilon = 0.5 h = 0.5 a/\mathcal{M}$.

\section{Results}
\label{sec:results}	

\begin{figure}
\epsscale{1.2}
\plotone{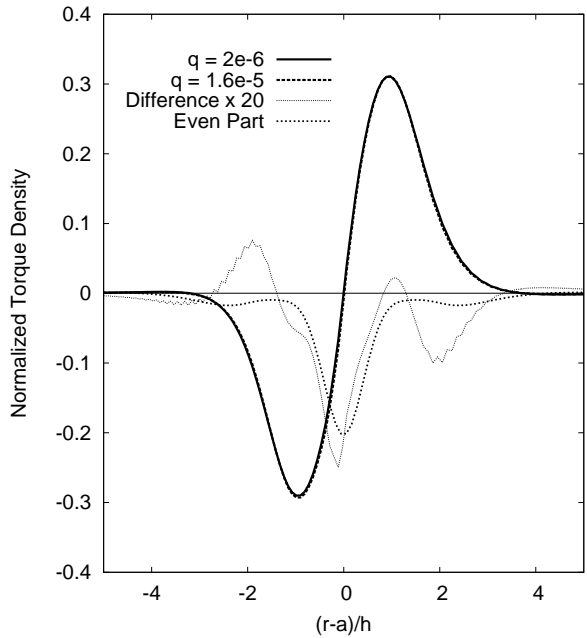}
\caption{ Departure from linear theory is measured in the torque density.  Torque densities (normalized by the factor $q^2 \mathcal{M}^4 \Sigma_0 a^3 \Omega^2$) are plotted for two planets, with mass ratios $q = 2 \times 10^{-6}$ and $1.6 \times 10^{-5}$.  The disk Mach number is 20 and the viscosity $\nu = 2.5 \times 10^{-5} a^2 \Omega_p$, corresponding to $\alpha = 0.01$ at the planet's orbital radius.  Since the torque density is properly normalized, these curves lie nearly exactly on top of one another.  The difference must be a nonlinear effect; this is also included in the figure (multiplied by 20 for visibility).  Taking the even part of this difference shows a strictly negative torque that the planet exerts on the gas, as it coaxes material past its orbit.  This negative torque on the disk implies a positive corotation torque on the planet.
\label{fig:tdens} }
\end{figure}

A total of $96$ systems are evolved for $5000$ orbits each.  The parameters explored are mass ratio $q$, disk Mach number $\mathcal{M}$ (or equivalently disk aspect ratio $h/r = 1/\mathcal{M}$), and dimensionless viscosity $\alpha = \nu \mathcal{M}^2 / a^2 \Omega_p$.  The $96$ systems include eight different disk models (choices of $\alpha$ and $\mathcal{M}$), and $12$ values of $q$ per disk model.  The canonical disk in this study has $\mathcal{M} = 20$, $\alpha = 10^{-2}$.  Figure \ref{fig:tdens} shows a plot of torque density in the canonical disk for two different planets, $q = 2 \times 10^{-6}$ and $q = 1.6 \times 10^{-5}$.  The torque density is normalized such that in the linear limit ($q \ll \mathcal{M}^{-3}$), it is independent of mass ratio.

Since both of these planets have a very nearly linear influence, their torque densities (after normalization) are nearly identical.  The difference between the two will reflect a nonlinear effect.  This difference (times 20) is also plotted in Figure \ref{fig:tdens}.  There are several contributions to this difference, but only the symmetric part will contribute to the net torque on the planet, since the antisymmetric part integrates to zero.  The even part of this difference is therefore also plotted, and this clearly represents a strictly negative torque that the planet exerts on the disk, which is interpreted here as the planet shepherding gas past it so as to maintain a uniform mass flux (see also \cite{2003ApJ...588..494M}).  Since the planet's nonlinear influence here is a strictly negative torque, there must be a positive backreaction torque on the planet, pushing it outward.

\begin{figure}
\epsscale{1.2}
\plotone{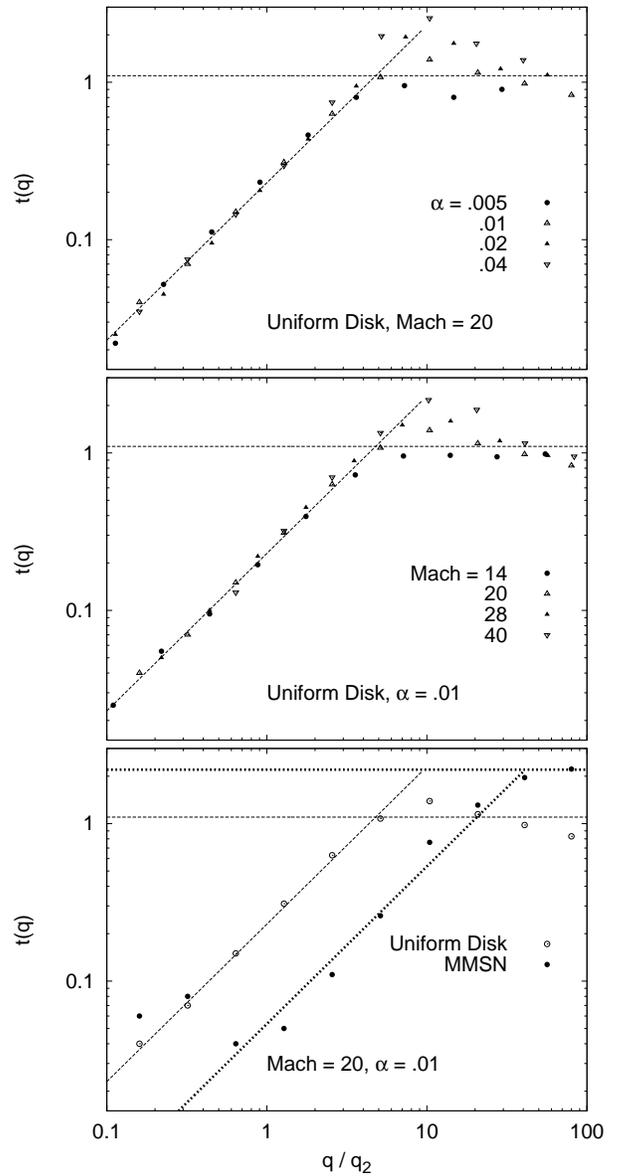}
\caption{ Scaling of $t(q)$ is measured for all $96$ systems by normalizing the torque and subtracting off a constant term.  For low-mass planets, $t(q) = B q/q_2$, where $B = 0.23$ and $q_2 \equiv \sqrt{\alpha}/\mathcal{M}^3$.  This is clearly shown in the figure; the low-mass results all lie on top of one another, and exhibit this linear scaling.  For higher-mass planets, $t(q)$ peaks and is then a decreasing function of $q$, exhibiting an additional scaling with $\alpha$ and $\mathcal{M}$.  The lower panel also shows a comparison between the uniform disk model, and the MMSN model with $\Sigma(r) \propto r^{-3/2}$.  The MMSN also exhibits an outward nonlinear corotation torque with a similar scaling in mass, but with a significantly smaller overall constant.  Horizontal lines in the figure denote the constant $A = 1.1$ (or $A = 2.2$ in the MMSN model).  Any time $t(q) > A$, this results in positive net torque, and outward migration.
\label{fig:torque} }
\end{figure}

Figure \ref{fig:torque} shows a measurement of $t(q)$ as defined by (\ref{eqn:tq}) for all $96$ systems.  $t(q)$ is simply calculated by dividing the total torque by $\Gamma_0$, and then subtracting off a constant.

The figure is consistent with a linear scaling of $t(q)$:

\begin{equation}
t(q) = B q/q_2,
\end{equation}

where $B$ is a constant independent of $q$, $\alpha$, or $\mathcal{M}$, and $q_2$ has the following scaling:

\begin{equation}
q_2 = \sqrt{\alpha} / \mathcal{M}^3.
\label{eqn:q2}
\end{equation}

Note that this is distinct from the characteristic mass ratio of nonlinearity $q_{NL} \equiv \mathcal{M}^{-3}$, the mass ratio at which the planet's Hill radius is of order a scale height.  In fact, for $\alpha \sim 0.01$, the corotation torque can be significant $q \sim q_2$ while still being only weakly nonlinear $q < q_{NL}$.  This scaling is obeyed for all $q \ll q_{NL}$, and for all choices of $\alpha$ and $\mathcal{M}$.  Equation (\ref{eqn:q2}) therefore appears to be a very robust result.

Moreover, this scaling is significantly distinct from the scaling predicted by \cite{2011MNRAS.410..293P}.  In that study, nonlinear torques were said to be strictly a function of a paramater called $p$, where

\begin{equation}
p \sim {\mathcal{M} \over \sqrt{\alpha}} (x_s/a)^{3/2}.
\end{equation}

Assuming $x_s \sim a \sqrt{ q \mathcal{M} }$, this gives

\begin{equation}
p \sim q^{3/4} \mathcal{M}^{7/4} \alpha^{-1/2} \sim (q/q_p)^{3/4},
\end{equation}

where $q_p \equiv \alpha^{2/3} / \mathcal{M}^{7/3}$, which scales differently from the empirically found $q_2 = \sqrt{\alpha} / \mathcal{M}^3$.  The scalings of \cite{2011MNRAS.410..293P} are derived from theoretical results, and only the viscosity was varied in their numerical simulations.  The explanation for why these theoretical results diverge from the study presented here should be investigated in future work.

However, this scaling of $t(q)$ breaks down at some point, when $t(q)$ peaks (Figure \ref{fig:torque}).  In summary, the torque for low-mass (weakly nonlinear) planets can be well-approximated by the formula

\begin{equation}
\Gamma = \Gamma_0 ( -A + B q/q_2 ),~~~~ q<q_{\rm peak}
\label{eqn:21}
\end{equation}

where $A = 1.1$, $B = 0.23$, and $q_2 = \sqrt{\alpha}/\mathcal{M}^3$, and $q_{\rm peak}$ is the value of $q$ at which $t(q)$ attains its maximum, above which $t(q)$ is no longer equal to $B q/q_2$, and therefore equation (\ref{eqn:21}) is no longer applicable.  The constants $A$ and $B$ were obtained for a uniform disk model.  For a disk with $\Sigma \propto r^{-3/2}$ (MMSN), the obtained constants were $A = 2.2$ and $B = 0.053$, so that the nonlinear torques were not as significant.  However, torque reversal was still seen in the MMSN case, for Jupiter-mass planets.  Incidentally, the values of $A$ found in this study are consistent with formulas calculated in previous works, for low-mass planets in isothermal disks \citep[e.g.][]{2002ApJ...565.1257T, 2010ApJ...724..730D}.

The scaling of $q_{\rm peak}$ can be checked empirically.  Finding $q_{\rm peak}$ in the seven uniform disk models in this study, an approximate scaling is found:

\begin{equation}
q_{\rm peak} \approx C \alpha / \mathcal{M}^{2.3},
\end{equation}

where $C \approx 11$ for the uniform disk, and $C \approx 55$ for the MMSN.  It should be stressed, however, that this scaling and these constants are much less certain than the scaling of $q_2$, and the details of this should be more carefully probed in a future study.

The strongly nonlinear regime ($q > q_{\rm peak}$) departs from equation (\ref{eqn:21}), and the results are less clear.  Roughly speaking, many of the calculations exhibit a high-mass scaling close to $t(q) \propto q^{-1/4}$, but this scaling is much less robust than the weakly nonlinear case.  More detailed analysis of the strongly nonlinear regime is warranted, especially considering that $t(q)$ was found to be positive for all systems considered, which suggests either that viscous torques are not exerted on the planet or at least that the influence of viscous torques might not be significant for these systems.

\section{Discussion}
\label{sec:disc}

\begin{figure}
\epsscale{1.2}
\plotone{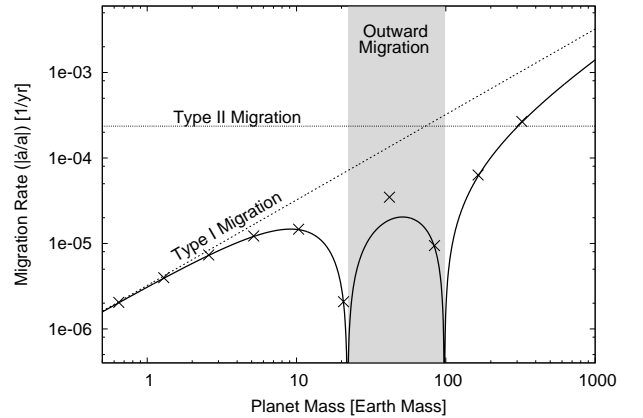}
\caption{ Typical disk parameters are chosen ($a = 1$ AU, $\Sigma_0 = 1700$ g/cm$^3$, $\mathcal{M} = 20$, $\alpha = 10^{-2}$, uniform disk model) and migration rates are calculated as a function of planet mass (assuming a solar-mass primary).  Crosses are direct numerical results, and the solid curve is the fitted scaling relation $\Gamma = \Gamma_0( -A + t(q) )$ (high-mass planets are fitted with $t(q) \propto q^{-1/4}$, though this scaling is much less certain).  Planets in the $20 - 100 M_{\earth}$ range migrate outward.  All planets with masses above $10 M_{\earth}$ migrate substantially slower than either predictions of Type I theory.
\label{fig:superearth} }
\end{figure}

Even in an isothermal disk, significant outward torques are present which can slow down or even reverse migration.  Figure \ref{fig:superearth} hilights the significance of this result.  Typical disk parameters ($\Sigma_0 = 1700$ g/cm$^3$, $\mathcal{M} = 20$, $\alpha = 10^{-2}$) are chosen to give a migration rate in inverse years at around 1 AU in the minimum mass solar nebula.  Also included are the Type I and Type II estimates \citep{1997Icar..126..261W}, which predict extremely fast inward migration for $10-100 M_{\earth}$ planets.  The nonlinear corotation torques result in an order-of-magnitude reduction in the migration rate for many of these systems, outward migration for planets in the $20-100 M_{\earth}$ range, and migration rates of zero for particular disk and planet parameters.  The migration rate of Jupiter in this figure happens to be just above the Type II rate, but this should not be interpreted as Type II migration, as the planet mass was kept on a fixed circular orbit, and therefore this study is insensitive to the disk-planet couplings that would be necessary to reproduce the Type II predictions.

\cite{2004ApJ...606..520M} pointed out that corotation torques are dependent on the density profile, and therefore Type I migration could be significantly reduced if $\Sigma(r)$ has the right slope.  This dependence on disk profile appears to be even more important for the nonlinear corotation torques; for example, the MMSN model ($\Sigma \propto r^{-3/2}$) only found outward migration for the Jupiter-mass planet (as seen in the bottom panel of Figure \ref{fig:torque}).

However, the question of whether a torque reversal occurs is also dependent on disk thickness and viscosity.  Some of the disk models considered exhibited no torque reversal at all.  As can be seen in Figure \ref{fig:torque}, the peak of $t(q)$ scales monotonically with $\alpha$ and $\mathcal{M}$.  Another way of stating this is that for low enough $\alpha$ or low enough $\mathcal{M}$, $q_{\rm peak} \ll q_2$, so that nonlinear torques peak before becoming comparable to Lindblad torques.  Thus, for low enough viscosity or a thick enough disk, no torque reversal will be seen.  In particular, the disk with $\alpha = 0.005$ and the disk with $\mathcal{M}=14$ both exhibited no torque reversal.  However, substantially thin disks with large enough $\alpha$ will have torque reversal for some range of planet masses (though this is again sensitive to the slope of the surface density).  Thus, there are many regions of parameter space in which one would not find torque reversal.  In particular, in the inviscid limit $\alpha \rightarrow 0$ one would not expect to find torque reversal, assuming that these observed trends continue in this limit.

To point out a specific example of a study which avoided this region of parameter space, \cite{2003ApJ...588..494M} chose a single viscosity, a single Mach number and a single planet mass for the majority of their calculations.  This study chose a $\sim 100 M_\earth$ planet, and as the density profile was $\Sigma(r) \propto r^{-3/2}$, this mass was below the threshold for outward migration, according to the calculations presented here.  In contrast, the current study varies planet mass, viscosity, and Mach number.  For a more recent example, the study of \cite{2014arXiv1411.3190D} chose typically low viscosities ($\alpha = 0.003$) and fixed $\mathcal{M} = 20$.  The one example of large enough viscosity $\alpha = 0.01$ was run with a single planet mass, so it is possible that this study, too, overlooked the regime outward migration (it should be noted that the regime of outward migration was not the focus of either of those works).  The much more thorough parameter space survey of \cite{2006ApJ...652..730M} found torque reversal in an isothermal disk, in agreement with the results presented here.

\begin{figure}
\epsscale{1.2}
\plotone{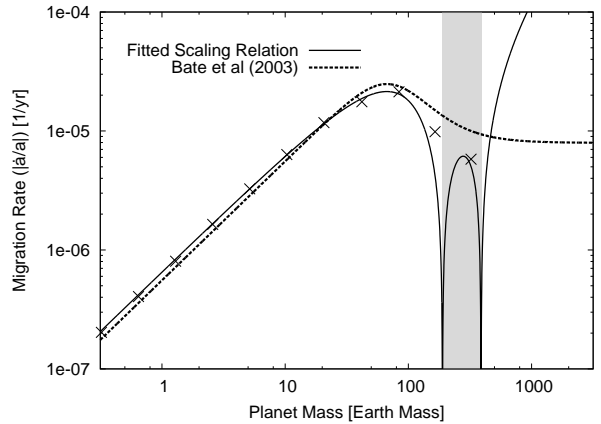}
\caption{ One potential misinterpretation of the data: torques in the MMSN disk model are scaled to match (as closely as possible) the parameters used in the study by \cite{2003MNRAS.341..213B} (see their Figure 11).  This is compared with the fitting formula provided by Bate et al., which strongly suggests that both studies are seeing the same phenomenon.  However, the interpretation by Bate et al. does not make sense for a planet kept on a fixed circular orbit, as was the case in both studies.  Moreover, this interpretation does not explain the outwardly-migrating Jupiter found for this disk model.
\label{fig:bate} }
\end{figure}

There have been many examples in which outward torques from partial gaps in isothermal disks have been observed in previous studies, with varying interpretations.  \cite{2003MNRAS.341..213B} studied torque as a function of planet mass and found a very similar curve to the one found in this study, but attributed it to the transition from Type I to Type II migration (See their Figure 11).  Figure \ref{fig:bate} shows the results found in the MMSN model in this study, scaled to match as closely as possible the results of \cite{2003MNRAS.341..213B} ($\Sigma_0 = $75 g/cm$^2$, $a = 5.2$ AU).  Also included is the fit produced by Bate et al. using the formula for transition from Type I to Type II migration \citep{1997Icar..126..261W}.  The differences between our results are very small, given the discrepancies between our disk models (Bate et al. studied a 3D disk which was not globally isothermal, and had surface density which scaled as $\Sigma(r) \propto r^{-1/2}$).  It seems possible that both studies are seeing the same phenomenon, but with a different interpretation.  However, since the planet is fixed at a given radius in both studies, the interpretation of \cite{2003MNRAS.341..213B} is very unlikely.  Even if one were to ignore the outward migration for Jupiter, the idea that the data reflect a transition to Type II is an unlikely interpretation, since these data can be re-scaled by our choice of $\Sigma_0$, due to the scale invariance of the hydro equations.  The fact that the migration rate appears to coincide with the viscous rate for large planets is entirely dependent on the choice of $\Sigma_0$, which is arbitrary.  It seems likely that Bate et al. did not see a transition to Type II migration, but might have in fact witnessed a transition toward outward migration (or at least suppressed migration due to nonlinear corotation torques).  It should be pointed out that \cite{2003MNRAS.341..213B} did not have data beyond a Jupiter mass, and this transition to ``Type II" might simply be interpreted in that paper to mean a transition from ``no gap" to ``gap", not necessarily a transition to migration at the viscous rate.

If torque reversal occurs, this will happen when the first and second terms of Equation (\ref{eqn:six}) are comparable,  when $A \approx B q/q_2$ (This is an approximate relation and not exact because higher-order terms should also be important in Equation (\ref{eqn:six})).  This implies a critical mass for the planet:

\begin{equation}
q_{crit} \approx (A/B) q_2 = 42 \sqrt{\alpha} \mathcal{M}^{-3}
\label{eqn:qcrit}
\end{equation}

($A/B \approx 42$ is appropriate for the MMSN model in this study, Figure \ref{fig:bate}).  When applying this formula, however, one must remember that torque reversal is not expected to occur at all if $q_{\rm crit} < q_{\rm peak}$.  Assuming $\alpha = 0.01$ and $\mathcal{M} = 20$,

\begin{equation}
m_{crit} \approx m_J.
\end{equation}

This transition to outward migration near a Jupiter mass is seen in Figure \ref{fig:bate}.  This torque reversal at a critical mass was also seen by \cite{2013ApJ...769...41D} (their Figure 3).  In that study, torque reversal was specifically seen for low viscosities, and the critical planet mass was found to scale as $q_{crit} \sim 70 \sqrt{\alpha} \mathcal{M}^{-3}$, consistent with the approximate formula (\ref{eqn:qcrit}) (scaling with Mach number was not measured in that paper).

In the study of \cite{2006ApJ...652..730M}, torque reversal was found in a uniform disk with $\nu = 10^{-5} a^2 \Omega_p$, $\mathcal{M} = 20$.  This is consistent with $q_2 = 7.9 \times 10^{-6}$, so that $q_{\rm crit} = 3.8 \times 10^{-5}$, which is precisely where that study found torque reversal to occur (see their Figure 3).  Note for this case that one can calculate $q_{\rm peak} \approx 4.5 \times 10^{-5}$, which is very close to $q_{\rm crit}$, so this is a marginal case where the nonlinear torque is just strong enough to change the sign of the total torque (note that the disk model used in Figure \ref{fig:superearth} has a higher viscosity $\nu = 2.5 \times 10^{-5} a^2 \Omega_p$ than Figure 3 of \cite{2006ApJ...652..730M}, but otherwise the disk models are essentially the same.  This is why, for example, the transition mass $q_{\rm crit}$ in Figure \ref{fig:superearth} is larger than in the study of \cite{2006ApJ...652..730M}).

Alternatively, if the mass is fixed, the torque can be reversed at a critical viscosity

\begin{equation}
\alpha_{crit} \approx ( (B/A) q \mathcal{M}^3 )^2.
\end{equation}

It should also be noted that these outward torques may be even more significant in three dimensions (3D), because in 3D the nonlinear corotation torques are likely less dependent on the disk's vortensity profile, since vortensity is not conserved.  Strong outward torques for Super-Earths in isothermal disks are already being seen in some of the parameter space of current 3D studies \citep{jeffrey}.

\acknowledgments

Resources supporting this work were provided by the NASA High-End Computing (HEC) Program through the NASA Advanced Supercomputing (NAS) Division at Ames Research Center.  I am grateful to Robin Dong, Eugene Chiang, Eliot Quataert, Sijme-Jan Paardekooper, Alessandro Morbidelli, and Matthew Bate for helpful comments and discussions.  I am especially grateful to Jeffrey Fung for confirming some of these results using the excellent PEnGUIn code.  I would also like to thank the anonymous referee for a very thorough review, which has vastly improved the clarity of this paper.

\begin{appendix}

\begin{figure}
\epsscale{1.2}
\plotone{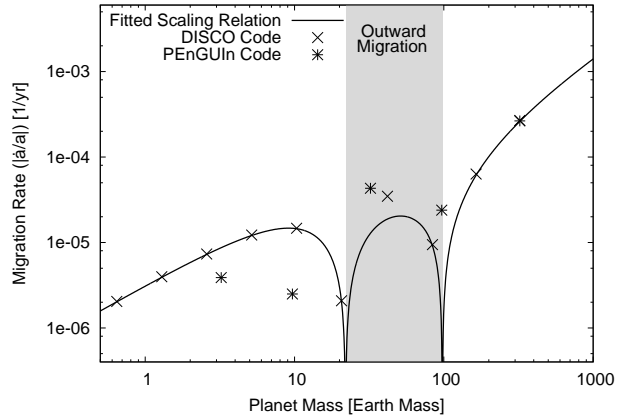}
\caption{ The same overall picture is captured with two different codes, DISCO (the main code used in this study) and the PEnGUIn code \citep{2014ApJ...782...88F}.  The codes used nearly identical numerical setups, except that the PEnGUIn code had a somewhat larger inner boundary radius, and was run for a shorter time ($1000$ orbits instead of $5000$ orbits).  Both codes find outward migration in the shaded $20 - 100 M_{\earth}$ range of planetary masses for the given disk model.  PEnGUIn finds larger outward torques, but is otherwise completely consistent with the results of DISCO. 
\label{fig:jeffrey} }
\end{figure}

\section{Independent Code Validation}

The results of this work are potentially important for answering the question of how to halt the migration of mid-range planets.  Therefore, it is worthwhile to validate these results with independent numerical calculations using a different hydro code.

The results in Figure \ref{fig:superearth} above have been checked independently using the GPU-enhanced PEnGUIn code, courtesy of Jeffrey Fung \citep{2014ApJ...782...88F, 2014ApJ...790...78F}.  PEnGUIn uses a fixed grid in the corotating frame, with third-order piecewise parabolic reconstruction, and integrates a different analytic form of the hydro equations from DISCO.  The architecture and numerical methods of these two codes are very distinct from one another, and therefore this represents a significantly independent check on the results of this work.

The numerical set-up is nearly identical to the parameters employed in this study, except that the inner boundary is located at $r_{\rm min} = 0.4 a$ instead of $0.25 a$, and the calculation was integrated for only $1000$ orbits, instead of $5000$.  The comparison of  results from the two codes is presented in Figure \ref{fig:jeffrey}.

The PEnGUIn code found larger positive nonlinear torques than DISCO, so that inward migration is somewhat slower and outward migration is slightly faster, but otherwise this represents a strong confirmation of the basic results found in this study.  In particular, PEnGUIn found outward migration in the same range of planet masses as DISCO (PEnGUIn's two data points in the shaded region of Figure \ref{fig:jeffrey} had positive net torque, while all others had negative net torque).

\end{appendix}



\begin{thebibliography}{}
\expandafter\ifx\csname natexlab\endcsname\relax\def\natexlab#1{#1}\fi

\bibitem[{{Bate} {et~al.}(2003){Bate}, {Lubow}, {Ogilvie}, \&
  {Miller}}]{2003MNRAS.341..213B}
{Bate}, M.~R., {Lubow}, S.~H., {Ogilvie}, G.~I., \& {Miller}, K.~A. 2003,
  \mnras, 341, 213

\bibitem[{{Bitsch} {et~al.}(2013){Bitsch}, {Crida}, {Morbidelli}, {Kley}, \&
  {Dobbs-Dixon}}]{2013AnA...549A.124B}
{Bitsch}, B., {Crida}, A., {Morbidelli}, A., {Kley}, W., \& {Dobbs-Dixon}, I.
  2013, \aap, 549, A124

\bibitem[{{Crida} \& {Morbidelli}(2007)}]{2007MNRAS.377.1324C}
{Crida}, A., \& {Morbidelli}, A. 2007, \mnras, 377, 1324

\bibitem[{{Crida} {et~al.}(2007){Crida}, {Morbidelli}, \&
  {Masset}}]{2007AnA...461.1173C}
{Crida}, A., {Morbidelli}, A., \& {Masset}, F. 2007, \aap, 461, 1173

\bibitem[{{D'Angelo} \& {Lubow}(2010)}]{2010ApJ...724..730D}
{D'Angelo}, G., \& {Lubow}, S.~H. 2010, \apj, 724, 730

\bibitem[{{Duffell} {et~al.}(2014){Duffell}, {Haiman}, {MacFadyen}, {D'Orazio},
  \& {Farris}}]{2014ApJ...792L..10D}
{Duffell}, P.~C., {Haiman}, Z., {MacFadyen}, A.~I., {D'Orazio}, D.~J., \&
  {Farris}, B.~D. 2014, \apjl, 792, L10

\bibitem[{{Duffell} \& {MacFadyen}(2012)}]{2012ApJ...755....7D}
{Duffell}, P.~C., \& {MacFadyen}, A.~I. 2012, \apj, 755, 7

\bibitem[{{Duffell} \& {MacFadyen}(2013)}]{2013ApJ...769...41D}
---. 2013, \apj, 769, 41

\bibitem[{{D{\"u}rmann} \& {Kley}(2014)}]{2014arXiv1411.3190D}
{D{\"u}rmann}, C., \& {Kley}, W. 2014, ArXiv e-prints, arXiv:1411.3190

\bibitem[{{Edgar}(2007)}]{2007ApJ...663.1325E}
{Edgar}, R.~G. 2007, \apj, 663, 1325

\bibitem[{{Edgar}(2008)}]{2008arXiv0807.0625E}
---. 2008, ArXiv e-prints, arXiv:0807.0625

\bibitem[{{Fung} \& {Artymowicz}(2014)}]{2014ApJ...790...78F}
{Fung}, J., \& {Artymowicz}, P. 2014, \apj, 790, 78

\bibitem[{{Fung} {et~al.}(2014{\natexlab{a}}){Fung}, {Artymowicz}, \&
  {Wu}}]{jeffrey}
{Fung}, J., {Artymowicz}, P., \& {Wu}, Y. 2014{\natexlab{a}}, (in preparation)

\bibitem[{{Fung} {et~al.}(2014{\natexlab{b}}){Fung}, {Shi}, \&
  {Chiang}}]{2014ApJ...782...88F}
{Fung}, J., {Shi}, J.-M., \& {Chiang}, E. 2014{\natexlab{b}}, \apj, 782, 88

\bibitem[{{Goldreich} \& {Tremaine}(1980)}]{1980ApJ...241..425G}
{Goldreich}, P., \& {Tremaine}, S. 1980, \apj, 241, 425

\bibitem[{{Ida} \& {Lin}(2008)}]{2008ApJ...673..487I}
{Ida}, S., \& {Lin}, D.~N.~C. 2008, \apj, 673, 487

\bibitem[{{Kley}(1999)}]{1999MNRAS.303..696K}
{Kley}, W. 1999, \mnras, 303, 696

\bibitem[{{Kley} \& {Nelson}(2012)}]{2012ARAnA..50..211K}
{Kley}, W., \& {Nelson}, R.~P. 2012, \araa, 50, 211

\bibitem[{{Lin} \& {Papaloizou}(1986)}]{1986ApJ...309..846L}
{Lin}, D.~N.~C., \& {Papaloizou}, J. 1986, \apj, 309, 846

\bibitem[{{Lubow} \& {D'Angelo}(2006)}]{2006ApJ...641..526L}
{Lubow}, S.~H., \& {D'Angelo}, G. 2006, \apj, 641, 526

\bibitem[{{Lubow} {et~al.}(1999){Lubow}, {Seibert}, \&
  {Artymowicz}}]{1999ApJ...526.1001L}
{Lubow}, S.~H., {Seibert}, M., \& {Artymowicz}, P. 1999, \apj, 526, 1001

\bibitem[{{Masset} \& {Snellgrove}(2001)}]{2001MNRAS.320L..55M}
{Masset}, F., \& {Snellgrove}, M. 2001, \mnras, 320, L55

\bibitem[{{Masset}(2002)}]{2002AnA...387..605M}
{Masset}, F.~S. 2002, \aap, 387, 605

\bibitem[{{Masset} {et~al.}(2006){Masset}, {D'Angelo}, \&
  {Kley}}]{2006ApJ...652..730M}
{Masset}, F.~S., {D'Angelo}, G., \& {Kley}, W. 2006, \apj, 652, 730

\bibitem[{{Masset} \& {Papaloizou}(2003)}]{2003ApJ...588..494M}
{Masset}, F.~S., \& {Papaloizou}, J.~C.~B. 2003, \apj, 588, 494

\bibitem[{{Menou} \& {Goodman}(2004)}]{2004ApJ...606..520M}
{Menou}, K., \& {Goodman}, J. 2004, \apj, 606, 520

\bibitem[{{Mordasini} {et~al.}(2009{\natexlab{a}}){Mordasini}, {Alibert}, \&
  {Benz}}]{2009AnA...501.1139M}
{Mordasini}, C., {Alibert}, Y., \& {Benz}, W. 2009{\natexlab{a}}, \aap, 501,
  1139

\bibitem[{{Mordasini} {et~al.}(2009{\natexlab{b}}){Mordasini}, {Alibert},
  {Benz}, \& {Naef}}]{2009AnA...501.1161M}
{Mordasini}, C., {Alibert}, Y., {Benz}, W., \& {Naef}, D. 2009{\natexlab{b}},
  \aap, 501, 1161

\bibitem[{{Paardekooper} {et~al.}(2011){Paardekooper}, {Baruteau}, \&
  {Kley}}]{2011MNRAS.410..293P}
{Paardekooper}, S.-J., {Baruteau}, C., \& {Kley}, W. 2011, \mnras, 410, 293

\bibitem[{{Paardekooper} \& {Mellema}(2006)}]{2006AnA...459L..17P}
{Paardekooper}, S.-J., \& {Mellema}, G. 2006, \aap, 459, L17

\bibitem[{{Paardekooper} \& {Papaloizou}(2008)}]{2008AnA...485..877P}
{Paardekooper}, S.-J., \& {Papaloizou}, J.~C.~B. 2008, \aap, 485, 877

\bibitem[{{Tanaka} {et~al.}(2002){Tanaka}, {Takeuchi}, \&
  {Ward}}]{2002ApJ...565.1257T}
{Tanaka}, H., {Takeuchi}, T., \& {Ward}, W.~R. 2002, \apj, 565, 1257

\bibitem[{{Ward}(1986)}]{1986Icar...67..164W}
{Ward}, W.~R. 1986, \icarus, 67, 164

\bibitem[{{Ward}(1991)}]{1991LPI....22.1463W}
{Ward}, W.~R. 1991, in Lunar and Planetary Science Conference, Vol.~22, Lunar
  and Planetary Science Conference, 1463

\bibitem[{{Ward}(1997)}]{1997Icar..126..261W}
---. 1997, \icarus, 126, 261

\end{thebibliography}

\end{document}